\documentclass[runningheads,orivec]{llncs}
\usepackage[T1]{fontenc}
\usepackage{graphicx}

\usepackage{amsmath}

\usepackage{booktabs}

\begin{document}

\newcommand\blfootnote[1]{%
  \begingroup
  \renewcommand\thefootnote{}\footnote{#1}%
  \addtocounter{footnote}{-1}%
  \endgroup
}

\title{Multi-modal Liver Segmentation and Fibrosis Staging Using Real-world MRI Images}
\titlerunning{ }
%

\author{Yang Zhou\inst{1}\orcidID{0000-0001-5474-9605} \and
Kunhao Yuan\inst{2}\orcidID{0000-0002-9877-0219} \and
Ye Wei\inst{3}\orcidID{0009-0003-1842-5867} \and Jishizhan Chen\inst{1}\orcidID{0000-0002-1784-319X}}
\authorrunning{Y. Zhou et al.}

%
\institute{Multiscale X-ray Imaging (MXI) Lab, Department of Mechanical Engineering, University College London, London, UK  \\
\email{\{yang.zhou,jishizhan.chen\}@ucl.ac.uk} \and
Centre for Clinical Brain Sciences, University of Edinburgh, Edinburgh, UK \\
\email{kyuan3@ed.ac.uk} \and
MRC Weatherall Institute of Molecular Medicine, University of Oxford, Oxford, UK \\
\email{ye.wei@imm.ox.ac.uk}}
\maketitle              
\blfootnote{This paper has been accepted to the International Conference on Medical Image Computing and Computer-Assisted Intervention (MICCAI) 2025 CARE: Comprehensive Analysis \& Computing of Real-world Medical Images Challenge.}

\begin{abstract}
Liver fibrosis represents the accumulation of excessive extracellular matrix caused by sustained hepatic injury. It disrupts normal lobular architecture and function, increasing the chances of cirrhosis and liver failure. Precise staging of fibrosis for early diagnosis and intervention is often invasive, which carries risks and complications. To address this challenge, recent advances in artificial intelligence-based liver segmentation and fibrosis staging offer a non-invasive alternative. As a result, the CARE 2025 Challenge aimed for automated methods to quantify and analyse liver fibrosis in real-world scenarios, using multi-centre, multi-modal, and multi-phase MRI data. This challenge included tasks of precise liver segmentation (LiSeg) and fibrosis staging (LiFS). In this study, we developed an automated pipeline for both tasks across all the provided MRI modalities. This pipeline integrates pseudo-labelling based on multi-modal co-registration, liver segmentation using deep neural networks, and liver fibrosis staging based on shape, textural, appearance, and directional (STAD) features derived from segmentation masks and MRI images. By solely using the released data with limited annotations, our proposed pipeline demonstrated excellent generalisability for all MRI modalities, achieving top-tier performance across all competition subtasks. This approach provides a rapid and reproducible framework for quantitative MRI-based liver fibrosis assessment, supporting early diagnosis and clinical decision-making. Code is available at \url{https://github.com/YangForever/care2025\_liver\_biodreamer}. 

\end{abstract}
\section{Introduction}
Liver fibrosis is a key pathological process during chronic liver disease (CLD) that represents a critical global health challenge and leads to an estimated two million deaths annually and substantial socioeconomic costs~\cite{devarbhavi2023global}. Fibrosis can progressively disrupt liver microstructures and functions, potentially resulting in cirrhosis and even liver failure~\cite{gines2021liver}. Nevertheless, early fibrosis is often asymptomatic and potentially reversible with timely intervention~\cite{czaja2014prevention}, demanding the need for accurate and reproducible staging methods. The gold standard for fibrosis staging remains invasive liver biopsy \cite{kazi2024noninvasive}, but it is limited in clinical scalability. Magnetic resonance imaging (MRI) offers a promising non-invasive alternative due to its superior high soft tissue contrast. Multi-modal MRI further enriches analysis by providing complementary insights from different imaging sequences of the same anatomical structures.

Effective fibrosis staging on MRI is highly dependent on accurate liver segmentation. Although recent developments in artificial intelligence (AI), especially U-Net and its variants~\cite{wang2023deep,oh2023automated}, have greatly advanced medical segmentation tasks, clinical segmentation applications on MRI remain challenging due to its diverse sequences and protocols. Therefore, to integrate medical AI into real-world clinical practice, the Comprehensive Analysis\&computing of REal-world medical images (CARE2025) proposed a Challenge track, named CARE-Liver. The track aimed to facilitate Liver Segmentation (LiSeg) and Fibrosis Staging (LiFS) on multi-centre, multi-modal, and multi-phase MRI datasets with limited single-modal annotations. The data modality involved non-contrast data: T1-weighted imaging (T1WI), T2-weighted imaging (T2WI), and diffusion-weighted imaging (DWI), and contrast-enhanced data from different phases of Gd-EOB-DTPA (GED1-4).

To address the challenge of limited annotation in medical image segmentation, previous studies, such as in~\cite{wu2022minimizing}, have investigated semi-supervised learning (SSL) methods to propagate the segmentation capabilities to unlabelled data. Nonetheless, the effectiveness of these approaches can be affected by multi-modal data, which present different features even on the same organ. A crucial step in multi-modal segmentation is image registration, which enables the transfer of the manual annotations that are typically only available on a single modality \cite{zhang2021modality}. With the liver segmentation, recent research on fibrosis staging, like~\cite{gao2023reliable,liu2025merit}, demonstrated promising results on GED-enhanced MRI by investigating sub-region multi-view features to minimise background interference of the entire scan. In fibrosis, patients often exhibit liver surface irregularities, lobar edge retraction, and subtle internal nodules~\cite{aube2017liver}, with MRI scans typically diffuse and widespread across the liver~\cite{yu2025mri}. As a result, effective algorithms must capture not only local pixel intensity, but also capture the organ’s complex shapes, textures, and directional features.

Motivated by these findings, we proposed an automated pipeline including multi-modal MRI registration, multi-modal liver segmentation, and fibrosis staging across all modalities using the features from segmented areas. In summary, in this study, we (i) extended the GED4 training dataset using mix-cases data augmentation; (2) enabled multi-modal liver segmentation with generated pseudo-labels from GED4 annotations using a rigid co-registration method; (iii) applied nnUNet for liver segmentation across all the modalities provided in the track without using additional open-source datasets; (iv) proposed a fibrosis staging method using Random Forest classifiers based on STAD (Shape, Textural, Appearance, Directional) features derived from segmented regions-of-interest (ROIs). Our proposed method achieved top-tier performance in both LiSeg and LiFS tasks on the hold-out Validation dataset and demonstrated promising generalised capability on the hold-out out-of-distribution Test dataset.

\section{Dataset}

\begin{table}[t]
    \centering
    \caption{Data distribution of the Training dataset from the CARE-Liver 2025 track.}
    \label{tab:dataset}
    \begin{tabular*}{0.9\textwidth}{@{}c @{\extracolsep{\fill}}cccccc@{}}
        \toprule
        & & {\textbf{Annotated}} & \multicolumn{4}{c}{\textbf{Fibrosis Stage}} \\
        \cmidrule(l){4-7}
        \textbf{Vendor} & {\textbf{Cases}} & {\textbf{Cases (GED4)}} & {\textbf{S1}} & {\textbf{S2}} & {\textbf{S3}} & {\textbf{S4}} \\
        \midrule
        A               & 130              & 10                      & 55            & 64             & 12            & 38             \\
        B1              & 170              & 10                      & 39            & 29             & 11            & 91             \\
        B2              & 60               & 10                      & 3             & 10             & 9             & 38             \\
        \midrule
        \textbf{Total}  & \textbf{360}     & \textbf{30}             & \textbf{97}   & \textbf{64}    & \textbf{32}   & \textbf{167}   \\
        \bottomrule
    \end{tabular*}
\end{table}

The study cohort of the CARE-Liver track involved 610 patients diagnosed with liver fibrosis. 360 patients from three vendors, labelled as A, B1, and B2, were provided as a multi-phase and multi-centre Training dataset. The Training data statistics are shown in Table \ref{tab:dataset}. Each patient in the Training dataset has T2WI, DWI, and GED-enhanced dynamic MRIs. The GED-enhanced dynamic MRIs cover several key phases: non-contrast (T1WI), arterial (GED1), venous (GED2), delayed (GED3), and hepatobiliary (GED4).

However, it is challenging to perform tasks of LiSeg and LiFS due to the limited number of annotations and data imbalance. As shown in Table \ref{tab:dataset}, the annotations are only available for GED4 data, and the number of annotated cases is limited to 10 cases in each vendor, less than $10\%$ in total. Moreover, the distribution of cases from each vendor across the different fibrosis stages is significantly imbalanced, with 167 from S4, while only 32 from S3 in total. Additionally, MRI resolutions vary across different modalities, while the pre-alignments are not available, which makes sharing GED4 annotations directly difficult. Therefore, to expand the Training dataset, we applied data augmentation for the GED4 data and co-registration to other modalities, without utilising alternative open-source datasets. 

\subsubsection{Data Augmentation} 
We augmented annotated GED4 data by generating synthetic examples, using a structured vendor-wise instance mixing approach~\cite{zhang2023carvemix,Yuanprmix}. For each annotated GED4 sample as a source, we randomly selected the other five annotated samples within the same vendor as targets. Then, the source foreground was extracted and isotropically scaled to match the target foreground region spatially. At the end, the transformed source foreground replaced the target foreground. This method improved anatomical and intensity diversity while preserving realistic organ topology and sharp label boundaries. To avoid damaging the organ structure, we deliberately excluded other common augmentations like translation, rotation, Gaussian noise, and foreground-biased patch cropping.

\subsubsection{Multi-modal co-registration} 
To facilitate the multi-modal segmentation and fibrosis staging, we implemented rigid co-registration between manually annotated GED4 data and other modalities. The registration searched for a 3D Euler transformation along the translation, rotation, and scalling in the space, using mutual information (MI) as a metric optimised by a gradient descent algorithm. The registration was implemented utilising the SimpleITK Python package. Finally, the training dataset was expanded from the initial 30 annotated GED4 samples to a total of 354 multi-modal samples, incorporating both the original and augmented GED4 data. 

\section{Methods}
\begin{figure}[t]
    \centering
    \includegraphics[width=\textwidth]{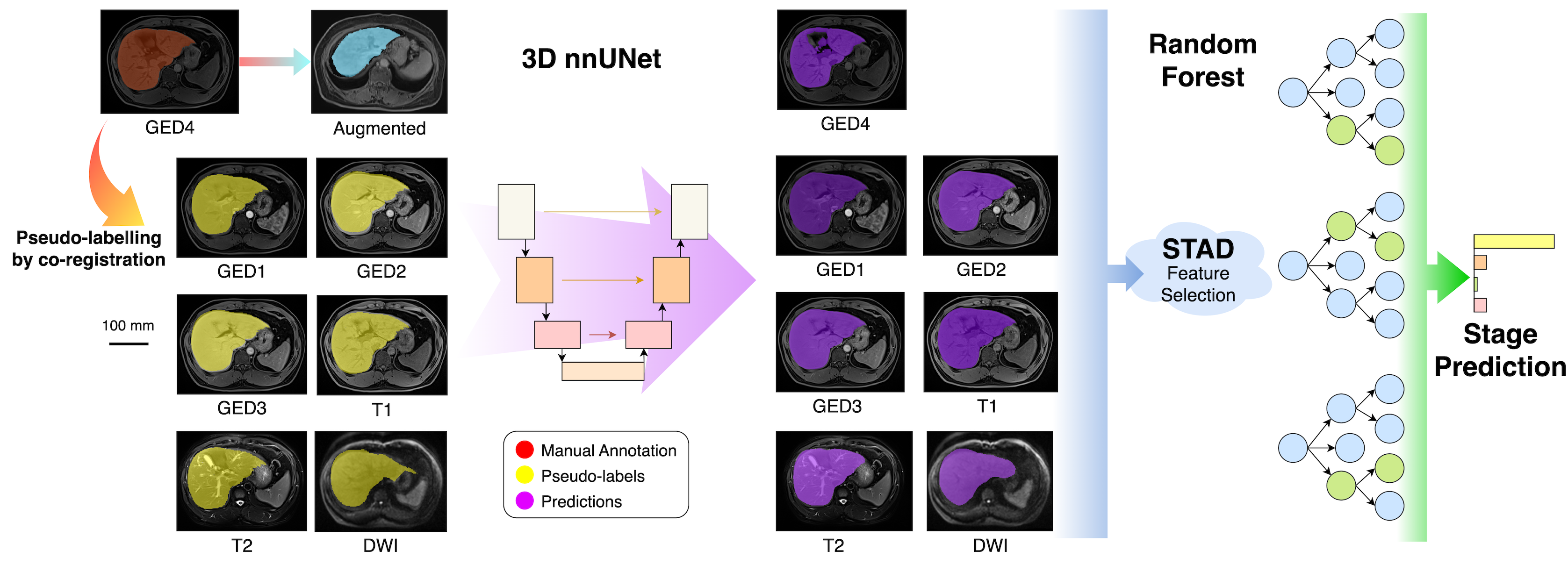}
    \caption{The automated pipeline for liver segmentation and fibrosis staging. A 3D nnUNet was trained on the dataset involving the original annotated GED4 data, augmented GED4 data, and the registered multi-modality data. After that, the fibrosis staging was performed by the Random Forest classifiers based on the STAD (Shape, Texture, Appearance, Directional) features selected from the segmentation predictions.}
    \label{fig:method}
\end{figure}

The track of CARE-Liver 2025 involves two tasks of automated liver segmentation (LiSeg) and Multi-phase fibrosis staging (LiFS). Each task is further divided into the non-contrasted subtask (T1WI, T2WI, and DWI) and the contrast-enhanced subtask (GED 1 to 4). Therefore, we proposed a pipeline that can process all the tasks, streamlining the LiSeg and LiFS. Without using other open-source datasets, as shown in Fig. \ref{fig:method}, we developed an automated method involving multi-modal liver segmentation and fibrosis staging based on the segmented liver regions. 

\subsection{LiSeg Task - Liver Segmentation}
On the registered multi-modal samples and augmented GED4 samples, we applied the state-of-the-art 3D nnUNet~\cite{isensee2021nnu} for liver segmentation. This method automates the biomedical image segmentation through network structure generation and training parameter selection based on the given training dataset. Before training the network, the dataset was pre-processed by an 8-bit conversion and data normalisation to prevent training bias, as the data types vary from each vendor and modality. 

The network was optimised by Stochastic Gradient Descent (SGD) with the weighted soft Dice loss and Cross-Entropy loss, shown as in Equation \ref{eq:seg_loss}.

\begin{equation}
\begin{split}
        L_{dc+ce} &= L_{soft\_Dice} + L_{Cross-Entropy} \\
            &=-w_{dc}\frac{2\sum{y_{pred}y_{true}}+f_{smooth}}{\sum{y_{pred}}+\sum{y_{true}}+f_{smooth}+\epsilon} - w_{ce}\sum{{y_{true}}\log{y_{pred}}} \,,
\end{split}
\label{eq:seg_loss}
\end{equation}
where the $y_{pred}$ denotes the predictions from the segmentation model and $y_{true}$ is the ground truths. $f_{smooth}$ is the smoothing factor for the Dice score, with a default value of 1. The term $\epsilon$, set to $1\times e^{-8}$ by default, is set for numerical stability to prevent division-by-zero error. The weights $w_{dc}$ and $w_{ce}$ determine the relative contribution of the Dice loss and Cross-Entropy loss to the total loss, and both have a default value of 1.  

\subsection{LiFS Task - Fibrosis Stage Classification}

The LiFS process began with feature selection, followed by stage classification using efficient Random Forest classifiers. The feature selection phase characterized the liver's macro-level morphologies and basic signal intensity within the segmented liver regions from the LiSeg task. Specifically, the features include: 

\begin{itemize}
\item	\textbf{Shape} features, such as volume, surface area, sphericity, and solidity, were computed to quantify geometric changes associated with fibrosis, like surface nodularity or parenchymal shrinking. Sphericity, for instance, measures how closely the liver's form resembles a perfect sphere, with deviations potentially indicating pathological alteration. 
\item   \textbf{Texture} features were extracted to capture more complex patterns indicative of fibrotic tissue. The magnitude of the image gradient is computed to quantify local intensity changes and "edginess" within the tissue, reflecting the degree of structural irregularity. Moreover, texture is analysed using Gray-Level Co-occurrence Matrices (GLCM) on three orthogonal planes to assess the spatial relationship of voxels, with features like contrast revealing the tissue's coarseness.
\item   First-order \textbf{appearance} features were calculated from the distribution of voxel intensities within the liver mask. These include statistical moments like mean, standard deviation, skewness, and kurtosis, which reflect the overall tissue density and heterogeneity.
\item	\textbf{Directional} feature analysis involved computing eigenvalues from the structure tensor and Hessian matrix. The features, such as coherence and anisotropy, probe the underlying microstructural arrangement of the tissue by measuring local orientation and curvature, which are critical for detecting the organized patterns of fibrotic bands.
\end{itemize}

In general, the primary benefit of this comprehensive STAD feature extraction is its ability to generate a rich, multi-faceted quantitative signature of the liver. By combining simple morphological metrics with sophisticated, higher-order textural and directional features, it created a holistic profile capable of capturing subtle pathophysiological changes. In addition, to account for potential data variation across vendors, we further incorporated a flag indicating the scanner vendor. 

These objective and explainable features are ideal for training robust Random Forest classifiers to accurately stage liver fibrosis without the risks associated with surgical biopsies. Ultimately, this approach paves the way for non-invasive diagnostics, enabling more frequent monitoring of disease progression and facilitating earlier detection and intervention.

\section{Results and Discussion}
In this section, we evaluated the model's performance on the hold-out Validation dataset and Test dataset from the CARE-Liver Track. Both datasets included samples from vendors, the same as the Training dataset, while the Test dataset introduced out-of-distribution data from a new vendor. For LiSeg and LiFS, we only trained and applied a single model for the tasks without ensembling.   

\begin{figure}[t]
    \centering
    \includegraphics[width=\textwidth]{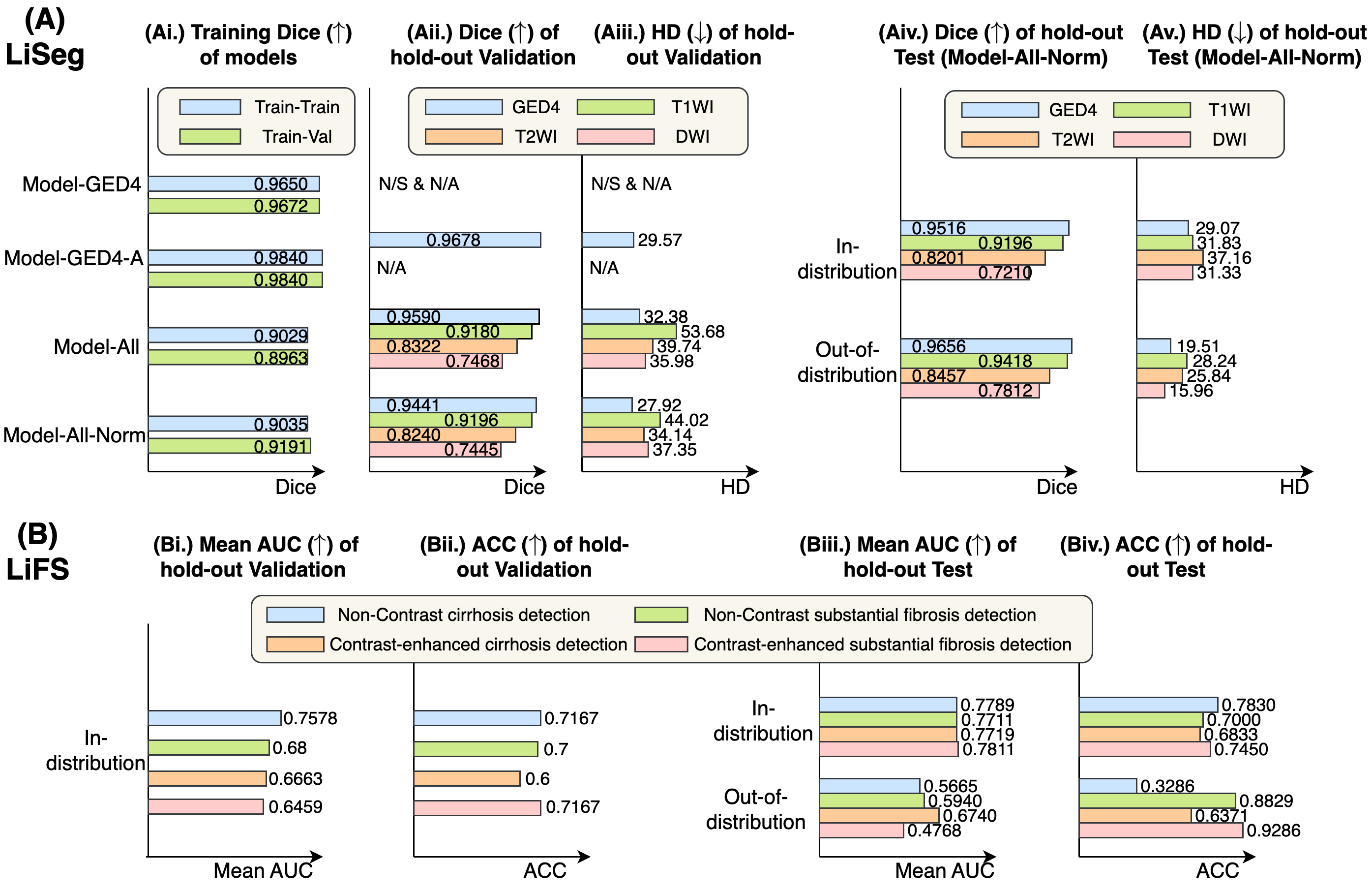}
    \caption{Quantitative evaluation of the method on LiSeg and LiFS. (A) shows the Dice and Hausdorff distance (HD) of different model configurations on (Ai.) the Training dataset and (Aii.)(Aiii.) the hold-out Validation dataset. (Aiv.)(Av.) illustrates the Model-All-Norm model performance on the hold-out Test dataset. (B) indicates the AUC and ACC of the Model-All on (Bi.)(Bii.)the hold-out Validation dataset and (Biii.)(Biv.) the hold-out Test dataset. (N/S denotes Not Submitted to the scoring system, and N/A is Not Applicable).}
    \label{fig:quantitative_evaluation}
\end{figure}

\subsection{Liver Segmentation}

LiSeg aimed for segmentation on non-contrast modalities, including T1WI, T2WI, and DWI, and segmentation on contrast-enhanced modality, GED4. To facilitate the liver segmentation on multi-modal MRI images, we explored multiple configurations to train a 3D nnUNet. Due to the restricted access to the Validation and Test datasets, the Training dataset was split as training (Train-Train) and validation (Train-Val) based on a ratio of 4:1. 

First, as shown in Fig. \ref{fig:quantitative_evaluation} (Ai.), we compared the models trained on annotated GED4 data only (Model-GED4) and on augmented data (Model-GED4-A), using identical training hyperparameters. The results show that Model-GED-A achieved better Dice scores on both Train-Train and Train-Val. 

Second, after multi-modal registration, we trained a model on all the data involving GED4, augmented GED4, and registered data, named Model-All. Due to the data varying from intensity ranges and data types, we further trained a model (Model-All-Norm) on normalised data, which are 8-bit range of 0-255. By evaluating the Dice scores on the Training and Validation datasets, both models achieved a similar performance of a range $\pm 0.02$ on all the MRI modalities. However, Model-All-Norm resulted in smaller Hausdorff Distances on GED4, T1WI, and T2WI. Therefore, we submitted the Model-All-Norm for the Test phase evaluation. It achieved better performance on Out-of-distribution data with $0.01\sim 0.05 $ Dice and $5\sim 15$ HD improvements compared to the performance on In-distribution data, showing promising generalised segmentation capability.  

\begin{figure}[t]
    \centering
    \includegraphics[width=\textwidth]{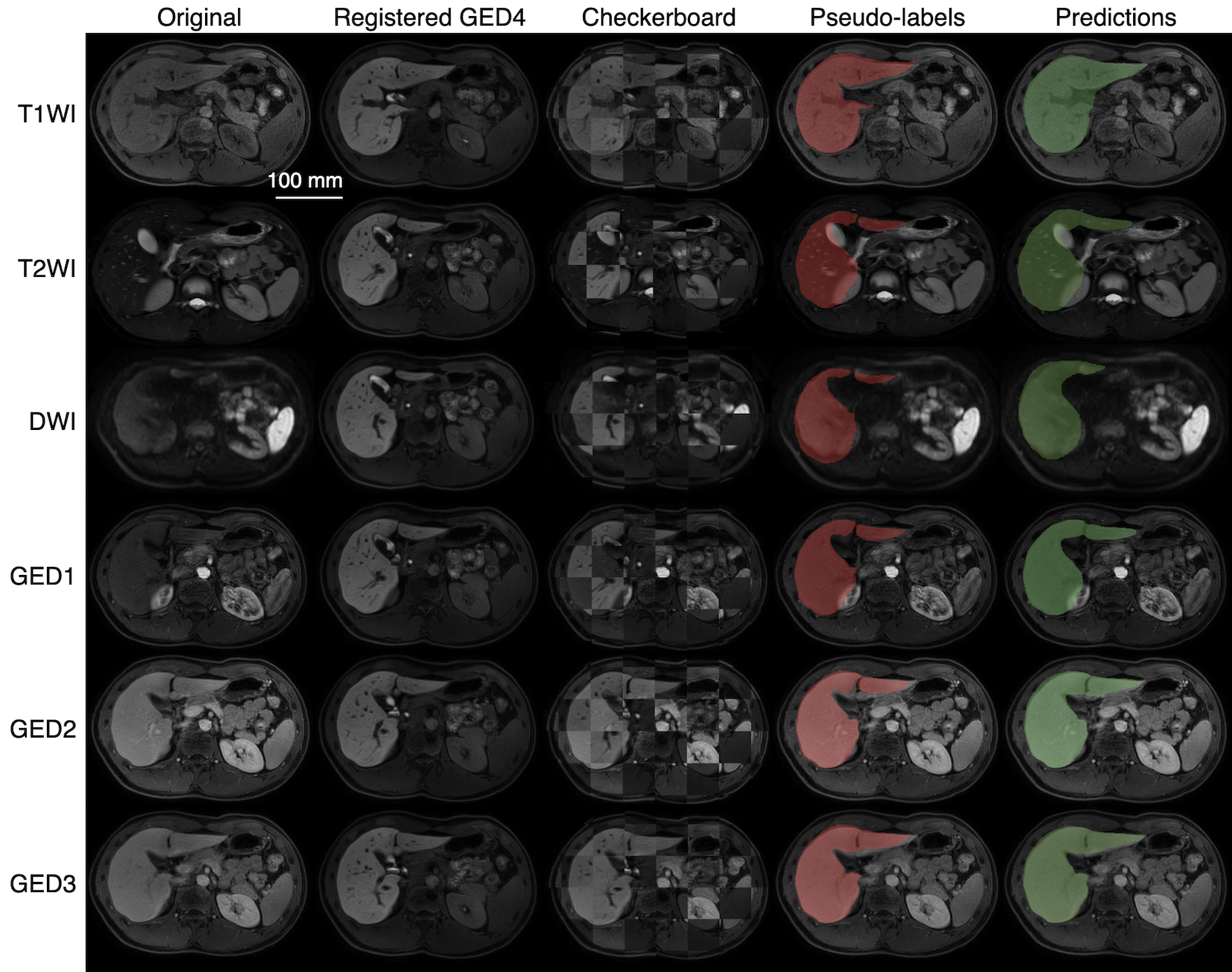}
    \caption{Results of co-registration, pseudo-labelling, and corresponding predictions. All the modalities are from a randomly selected case with the 2D slices taken from the middle of the MRI volume along the z-axis.}
    \label{fig:reg_seg_results}
\end{figure}

Figure \ref{fig:reg_seg_results} shows the results of co-registration and the predictions from Model-All-Norm. In this example, rigid co-registration achieved good alignment for modalities with contrast similar to GED4, including T1WI, GED2, and GED3. However, the alignment was slightly shifted for the remaining modalities. Notably, T2WI and DWI demonstrate low contrast on the liver region and show different liver features, which makes co-registration and their pseudo-labels inaccurate. This introduced bias during model training, causing lower Dice on both the hold-out Validation and Test datasets, with $\sim 0.12$ to $0.13$ decrease compared to GED4 segmentation, while only $\sim 0.05$ decrease for T1WI.   

\subsection{Liver Fibrosis Staging}

\begin{figure}[t]
    \centering
    \includegraphics[width=\textwidth]{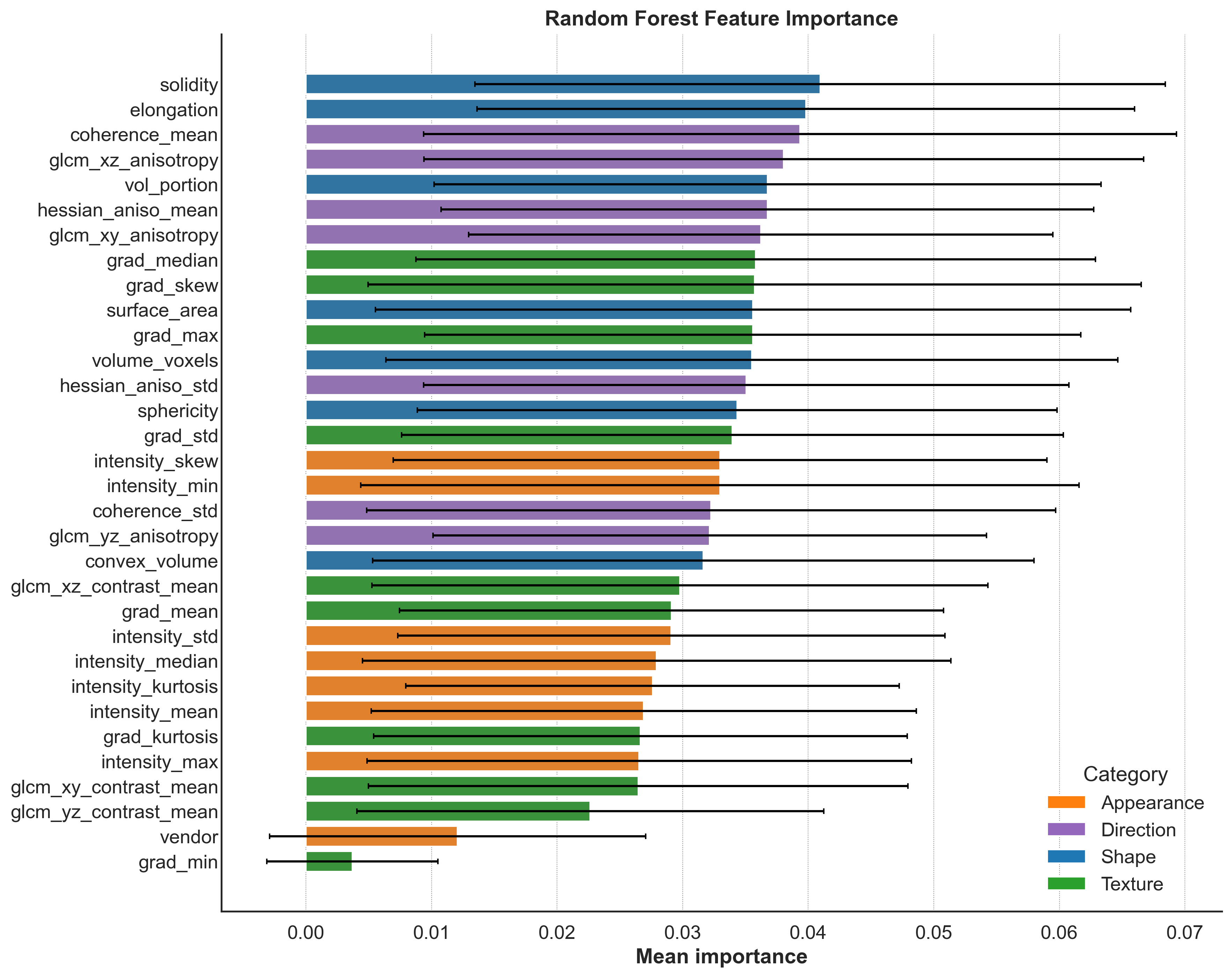}
    \caption{Importance analysis of the features selected for Random Forest classifiers. Standard deviation is shown as the error bar in black.}
    \label{fig:feature_importance}
\end{figure}

The LiFS aimed for two subtasks of cirrhosis detection (stage S4 vs. S1–S3) and substantial fibrosis detection (Stages S2–S4 vs. S1), evaluated separately on non-contrast and contrast-enhanced modalities. Using liver segmentation predictions on the Training dataset from Model-All, we isolated liver regions across various MRI modalities. From those regions, we extracted 32 STAD features, as illustrated in Fig. \ref{fig:feature_importance}, and applied Random Forest classifiers on a train-validation split ratio of 4:1. 

Fig.~\ref{fig:quantitative_evaluation} (B) demonstrates robust and consistent performance of our proposed model across both the hold-out Validation and in-distribution Test (which is also from vendors as Validation) datasets. Specifically, both metrics of mean Area Under the Curve (AUC) and Accuracy (ACC) improved on the Test dataset compared to those achieved in the Validation dataset. For example, the task of Contrast-enhanced substantial fibrosis detection increased approximately 0.14 of the mean AUC, while the Contrast-enhanced cirrhosis detection increased nearly 0.09 on ACC.

When our model was presented with the out-of-distribution (OOD) Test dataset, the detection performance was impacted with decreases in the mean AUC metric among both modalities and tasks. Particularly, there was a drop to 0.4768 in contrast-enhanced substantial fibrosis detection compared to 0.7811 achieved in the in-distribution Test dataset. This highlighted the challenge of domain shift that often appears in medical image processing. However, the ACC of non-contrast and contrast-enhanced substantial fibrosis detection improved to 0.8829 and 0.9286, respectively, compared to an average of 0.7 and 0.72 on in-distribution data from Validation and Test. A decrease in mean AUC with an increase in ACC can indicate that some of the selected training features failed in OOD data. For example, variations in voxel intensity can adversely affect appearance-based features, potentially leading to detection errors, whereas other biologically consistent characteristics, like directional features, may provide compensation in some cases. 


Therefore, we investigated the importance of the features used in the Random Forest classifiers, as shown in Fig.~\ref{fig:feature_importance}. The features were ranked by their mean importance score, with an error bar showing as standard deviation. The analysis results indicated that the shape-based and directional features are significant in the multi-modal fibrosis detection, with solidity and elongation emerging as the top predictive characteristics. Directional features, such as coherence (coherence\_mean) and anisotropy (glcm\_xz\_anisotropy), also showed high importance, as they capture biological properties by assessing tissue orientation and curvature. Appearance-based features ranked lower in importance following some of the texture-based features. This feature importance analysis aligned with our fibrosis detection results on the hold-out Test dataset, further demonstrating that the feature selection process can be improved when considering the OOD data.  

\subsection{Discussion}
One limitation that affects the performance of our proposed pipeline is the accuracy of the co-registration process. Inaccurate pseudo-labels can introduce bias during the segmentation training phase, making the model training difficult and leading overfitting problem. The error potentially impacts the feature selection process for the fibrosis staging task. In this study, we found that upsampling low-resolution data to higher-resolution data results in higher registration errors. Therefore, we registered annotated GED4 to other modalities due to the lower resolutions of other modalities, such as DWI, to improve the pseudo-labelling quality. To improve the registration and performance of downstream tasks, future work can include non-rigid co-registration methods that consider the detailed deformation during the registration process. Additionally, as shown in Fig. \ref{fig:reg_seg_results}, the segmentation predictions can complementarily correct the pseudo-labels by filling the missing areas due to registration errors. Hence, fine-tuning the model on all Training dataset after performing an all-data inference might improve the segmentation performance, as in previous work \cite{zhang2025computing}. 

Apart from the cumulative error from the registration and segmentation process, the fibrosis staging performance was also limited by the features selected. By accessing the performance on the hold-out Test dataset and the feature importance analysis, biologically consistent intrinsic features that display in multiple modalities were more reliable. Those features can make the classifiers more robust and generalisable. Therefore, in the future, more shape-based and directional features will be investigated and included in training the model with a reduction of intensity-based appearance features. In addition, although the Random Forest classifier provided interpretable features, it may underperform compared to recent deep learning approaches. The observed drop in mean AUC on out-of-distribution data highlights limitations in generalization, which likely stems from the reliance on handcrafted features. Our study’s primary contribution lies in integrating multi-modal, real-world MRI data into a clinically interpretable framework, establishing a baseline for fibrosis staging in this context. Future work could explore neural network–based methods to improve generalization, particularly when larger annotated datasets become available.

\section{Conclusion}
In this study, we proposed an automated pipeline that integrates liver segmentation and fibrosis staging on multi-modal real-world clinical MRI images. Based on this pipeline, the nnUNet learned liver segmentation from multi-modal registered MRI images. The segmented regions, then, were used for STAD feature extraction and fibrosis stage classification through efficient Random Forest classifiers. The STAD features captured the fibrosis characteristics and features revealed from the real-world MRI images. The results demonstrated that our pipeline achieved top-tier performance on both LiSeg and LiFS tasks in CARE-Liver Track using limited data and annotations, without needing training on additional large open-source datasets. This work potentially presented novel insights for practical clinical MRI applications, particularly in scenarios involving multi-modality and multi-task learning under limited annotations.

\begin{credits}
\subsubsection{\ackname}
The authors acknowledge the use of the UCL Myriad High Performance Computing Facility (Myriad@UCL), and associated support services, in the completion of this work.

\subsubsection{\discintname}
The authors declare that they have no competing interests.
\end{credits}
%
%
%
\bibliographystyle{splncs04}
\bibliography{mybibliography}
\end{document}